\documentclass[12pt,a4paper,reqno,pdfoutput=1]{amsart}
\usepackage{xypic,a4wide,amsmath,amssymb,amsthm}
\usepackage[utf8x]{inputenc}
\usepackage[T1]{fontenc}

\usepackage{amssymb}
\usepackage{amsmath}

\usepackage[dvips]{graphicx}
\usepackage{array}

\title[\ ]{Quantum   Painlev\'e  II  Lax Pair and Quantum (Matrix) Analogues of Classical  Painlev\'e  II equation}

\author{ Muhammad Waseem,  Irfan Mahmood and Hira Sohail}

\email{mahirfan@yahoo.com} \urladdr{}

\begin{document}

\maketitle

\begin{abstract}
%% Text of abstract
 In this article, we present a new  quantum Painlev\'e II  Lax pair  which explicitly   involves the Planck constant $ \hbar $ and an arbitrary field variable $v$ so these two objects make this new pair different from Flaschka-Newell Painlev\'e II  Lax pair and that pair appears as particular case of our's pair which consolidates the Painlev\'e II equation from quantum mechanical point of view. It is shown that the compatibility of quantum Painlev\'e II  Lax pair simultaneously yields a   quantum Painlev\'e II equation and a quantum commutation relation between field variable $v $ and independent  variable $z$. We  manifest that with the different choices of arbitrary field variable  system reduces to its classical version,  matrix  Painlev\'e II equation  and derivative matrix  Painlev\'e II equation. Further, we construct the gauge equivalence of quantum Painlev\'e II  Lax pair whose compatibility condition gives rise to quantum p34 equation that involves $\hbar $ with power $+1$  which makes our system more quantized as compared to the existed one that carries $\hbar $ with power $+2$.
 \end{abstract}
\section{Introduction}

 The classical Painlev\'e II (PII) equation among  Painlev\'e  six equations \cite{PP}  is an  only one parametric  equation  therefore it can be regarded  as a primary object in the hierarchy of  Painlev\'e   equations for understanding   their  algebraic and geometrical aspects connected to  parametric values. Painlev\'e equations are also  regarded as completely integrable equations as mentioned by  
 \cite{r5, r11, r12}  that they admit  linear representations, possess Hamiltonian structures. and obeyed the Painlev\'e test.

In context of their representations and towards their integrability  some of the significant and  remarkable results are achieved on PII equation. Right off the bat  very initial  results concerning its solutions  were obtained   in \cite{YAB, VOR, HAR}  where it has been proved that the PII equation possesses rational solutions  for integer values of parameter $\alpha$ and expressed in terms of Yablonski Vorob’ev polynomials \cite{YAB, VOR}. Where as for the  half odd-integer  values of parameter $\alpha$  PII equation admits  Airy's type solutions \cite{HAR} and  also  owns the B$\ddot{a}$cklund transformation that relates its two different solutions  through the  parameter  $\alpha$.  Subsequently, Flaschka and Newell \cite{FN}   expressed the rational solutions of PII equation as the logarithmic derivatives of determinants with its linear representation in classical framework. In addition to that  Kajiwara and Ohta \cite{KO}  generalised  PII solutions in devisme polynomial determinant and as well as in terms of Hankel determinant. \\
 
With addition of above properties one of the  very interesting aspects of PII equation in theory of integrable systems is its appearance as ordinary reduction of some solitonic equation, for example it has been shown in \cite{r4, MA} it merges as dimension reduction  of the KdV equation. 
 \begin{equation}\label{CPII} 
 u^{''} = 2 u^{3}-zu +  \alpha,
  \end{equation} 
  At very beginning towards the  derivation of  various analogues of classical PII equation  (\ref{CPII})  and its representation  on more advanced surfaces, first time  it has been manifested  in \cite{OS}  that   its derivative matrix version connected to matrix mKdv equation yields the through the  dimensional reduction. Subsequently another  direct matrix (quantum) version was derived \cite{NH, NGR} for classical PII equation  (\ref{CPII}) with its partner equation  as P34 equation  that  induces Planck constant explicitly that  gives the sense of quantization of PII  equation which itself does not contain  the Planck constant.  Later on its  Non-commutative (NC) analogue connected to nonabelian Toda equation presented by Retakh and Rubtsove  \cite{7}   possessing anti-commutation relation  between field variable $u(z; \alpha)$ and independent variable $z$ but does not carry quantum relation in its expression  to consolidate it from quantum mechanical point of view. After that its Darboux solutions were derived  incorporating non-commutative Toda equations at $n=1$ as seed solutions \cite{MIRFAN, IM}  in terms of quasideterminants  \cite{GelRet}. \\
  \subsection{Highlights}
  In this article, we present a  new  quantum PII  Lax Pair which explicitly involves the Planck constant $\hbar$ and an arbitrary field variable $v$ so these two objects make this new pair different from Flaschka-Newell PII  Lax pair as is discussed in section $3$  and section $4$ with details. Further we show  that the compatibility of quantum PII  Lax Pair simultaneously yields a   non-commutative quantum PII equation and a quantum commutation relation between field variable $v $ and independent  variable $z$. it is also manifested with different choices of arbitrary field variable resulting  system reduces to  derivative matrix  PII equation  and to its classical analogue. Further, the gauge equivalence of quantum PII  Lax pair is derived whose compatibility condition gives rise to quantum P34 equation which reduces to its classical analogue under classical limit as  $\hbar \rightarrow 0 $. The main results we obtain from compatibility of  quantum PII  Lax Pair as is given in section $3$ are briefly discussed as below
  
  \begin{equation}\label{Result} 
\left\{
\begin{array}{lr}
   u^{''}=2u^3-\frac{1}{2}[z,u]_{+}+\alpha \; \; \; \; \; \; (QPII)\\
\frac{d}{dz}( zu-uz)=-\frac{i}{2} \hbar  u \\ 

\nu^{'''}=2\nu^2\nu^{'}+2\nu^{'}\nu^2+2\nu \nu^{'}\nu-\nu-x\nu^{'}+4[\nu,\nu^{''}]_{-}  \; \; \; \; \; \; (MQPII)\\
\mathbf{p}^{''} = -\frac{1}{2} \mathbf{p^{'} }\mathbf{p^{-1} }\mathbf{p^{'} } + 2\mathbf{p^{2} }- \frac{\delta^{2} }{2}\mathbf{p^{-1} } - (z-\frac{1}{2} \hbar  ) \mathbf{p } \; \; \; \; \;(QP34)
\end{array}
\right..
 \end{equation} 
   
In above results  (\ref{Result}) expression  (QPII) is obtained directly from the  newly presented quantum PII  Lax pair equipped with quantum commutation relation between independent varible $z$ and field variable $u$ further it is applied to get derivative matrix PII equation (MQPII).  The compatibility condition of gauge equivalence quantum PII  Lax pair  yields the quantum P34 equation (QP34)  where as same has been derived in   \cite{NH, NGR} through the non-abelain PII symmetric form  which differs only from ours  (QP34) by  involving  Planck constant $\hbar$  in quadratic form
 \begin{equation}\label{NG} 
\left\{
\begin{array}{lr}
   u^{''}=2u^3-zu+\alpha \; \; \; \; \; \; (QPII)\\
   \mathbf{p}^{''} = -\frac{1}{2} \mathbf{p^{'} }\mathbf{p^{-1} }\mathbf{p^{'} } + 2\mathbf{p^{2} }- \frac{\delta^{2} }{2}\mathbf{p^{-1} } - (z-\hbar^2 ) \mathbf{p } \; \; \; \; \;(QP34)
\end{array}
\right.,
 \end{equation} 
and also does carry explicit quantum commutation relation  between field variable and independent variable.  Let make a explicit comparison between my calculated results  with results presented in  \cite{NH, NGR}  quantization of classical Painlev\'e II  equation with its  P34   partner equation. We present quantum PII Lax pair (see section 3) that directly involves the Planck constant $\hbar$ and an arbitrary field variable $v$. Its  compatibility condition produces quantum Painlev\'e II equation with a quantum commutation relation between field variable $v $ and independent  variable $z$ simultaneously where as its gauge equivalent pair generates quantum p34 equation that involves Planck constant $\hbar$  which involves Planck constant with power $+1$  that rescues it to be negligible as compare to  $ \hbar^{2} $  appears in  (\ref{NG})  presented by   \cite{NH, NGR}. The presence of Planck constant explicitly  with power $+1$  as $ \hbar $ in    Painlev\'e II  equation and in P34 equation  with qunatum commutation relation $\frac{d}{dz}( zu-uz)=-\frac{i}{2} \hbar  u$  (\ref{Result}) gives more better sense of its quantization and consolidates its from quantum mechanical point of view. Further our system also gives rise to the derivative matrix PII equation that seems similar to the derivative matrix PII equation  derived by Olver and Sokolov   \cite{OS} as the dimensional reduction of matrix mKdv equation (discussed as  section 3).

 \section{Different analogues of classical PII equation}
  This section encloses a brief  review on various analogues of Classical PII equation as its matrix  and non-commutative   versions.
    \subsection{Classical PII equation}
    The classical PII equation  (\ref{CPII}) initially  was proposed by P. Painlev\'e appeared as one of the member of six Painlev\'e  equations whose solutions possess parametric dependence expect PI equation, here classical means field variable $u(z; \alpha)$ and variable $z$ are scalars and commuting. 
The classical PII equation is integrable as its possesses linear representation \cite{FN}  and arises from the compatibility of following linear system
\begin{equation}\label{LPII} 
\Psi_{z}=U(z;\lambda)\Psi  , \hspace{0.5cm} \Psi_{\lambda}=V(z;\lambda)\Psi,  
\end{equation}

 with matrices $U$ and $V$  as
\begin{equation}\label{GLP} 
\left\{
\begin{array}{lr}
U =  -i \lambda \sigma_{3}  + u \sigma_{1}  \\
A= -i (4 \lambda^{2} +z + 2u^{2} ) \sigma_{3} +(4 \lambda u - \frac{\alpha}{\lambda} ) \sigma_{1} - 2v \sigma_{2}
\end{array}
\right.,
 \end{equation}
 here $\Psi$ is arbitrary two component column vector and    $\sigma_{j} $ are the Pauli spin matrices,
  $\sigma_{1} =  \begin{pmatrix}
   0 & 1 \\
   1 & 0
\end{pmatrix}$, 
   $\sigma_{2} = \begin{pmatrix}
0 & -i \\
i & 0
\end{pmatrix} $,   $\sigma_{3} = \begin{pmatrix}
1 & 0 \\
0 & -1
\end{pmatrix},$
where the matrices $U$ and $V$ are called the Flaschka-Newell Lax pair. \\
\textbf{Remark 2.1} The compatibility of gauge equivalent Flaschka-Newell Lax pair 
\begin{equation}
    \frac{\partial \Psi}{\partial \eta}=A \Psi,\hspace{1cm}\frac{\partial \Psi}{\partial z}=B \Psi,
\end{equation}
with matrices
\begin{equation}
    A=\begin{pmatrix}
        2u+\frac{\alpha+1/2}{2\eta}& 2i \eta+i q\\
        2i+\frac{i \sigma}{\eta}&-2u-\frac{\alpha+1/2}{2\eta}
    \end{pmatrix},
\end{equation}
\begin{equation}
    B=\begin{pmatrix}
        u&i\eta \\
        i &-u
    \end{pmatrix},
\end{equation}
 gives rise to PII symmetric form as below
\begin{equation}\label{csp2} 
\left\{
\begin{array}{lr}
 q^{'}= 2qu -\alpha+\frac{1}{2} \\
 r^{'}= -2ru + \alpha+\frac{1}{2}\\
u^{'}= \frac{1}{2}(q -r)
\end{array}
\right.,
 \end{equation} 
 where $\eta=\lambda^2$ and  matrices $A$ and $B$ are the gauge equivalence   \cite{AAAV} of Flschka-Newel Lax pair. On eliminating $u$ this can be shown that $r$ and $q$ satisfy P34 equation respectively. 
 \begin{equation}\label{q34} 
 r_{zz}=   \frac{r^{2}_{z}}{r}+2r^{2}-zr- \frac{1}{2r}(\alpha +  \frac{1}{2} )^{2},
\end{equation}
and 
 \begin{equation}\label{p34} 
q_{zz}=   \frac{q^{2}_{z}}{q}+2q^{2}-zq- \frac{1}{2q}(\alpha -  \frac{1}{2} )^{2},
\end{equation} 
where $u$ satisfies classical PII equation.
 \subsection{Derivative Matrix PII equation}
In theory of integrable systems it has been shown  that Painlev\'e  equations are obtained as  symmetry reductions  of various  integrable systems.  In context of their integrable symmetric reduction at the right of bat  its has been shown by Olver and Sokolov \cite{OS} that the ordinary reduction of Matrix mKdV equation
\begin{equation}
    v_t=v_{xxx}+3[v,v_{xx}]_{-}-6vv_xv,
\end{equation}
with transformation
\begin{equation}
    v(x,t)=u(z=xt^{-1/3})t^{-1/3},
\end{equation}
gives rise to ODE as follow
\begin{equation}
    u^{'''}=3u^{''}u-3uu^{''}+6uu^{'}u-\frac{1}{3}u-\frac{1}{3}zu^{'},
\end{equation}
which is derivative matrix PII equation that reduces to its classical analogue in scalar case.  
   \subsection{Matrix PII equation}
   The  Matrix (Quantum) analogue  of the classical P II equation derived in  \cite{NH, NGR}   with the help of non-abelian form P II  symmetric form (\ref{csp2})  as below
   \begin{equation}\label{spII} 
\left\{
\begin{array}{lr}
 q^{'}= uq+ qu -\alpha+\frac{1}{2} \\
 r^{'}= -ru-ur + \alpha+\frac{1}{2}\\
u^{'}= \frac{1}{2}( q -r)
\end{array}
\right.,
 \end{equation} 
 where $u$ satisfies matrix  Painlev\'e II equation
  $ u^{''}=2u^3-zu+\alpha_1-\alpha_0$ 
 and $q$, $r$ are the solutions of P34 equation which explicitly involve Planck constant, for example for $q$  the quantum P34 equation  can be obtained as follow
\begin{equation}
    q^{''}=\frac{1}{2}q^{'}q^{-1}q^{'}-4q^2+2zq-\frac{1}{2}(\alpha_1^2-\hbar^2)q^{-1},
\end{equation}
where the fields in non-abalian Painlev\'e II  symmetric form are subjected to quantum commutation relations as 
  \begin{equation}\label{qcr} 
[r,q]_{-}=2 \hbar u, \hspace{0.5cm} [u,q]_{-}=   [u,r]_{-}=\hbar ,
 \end{equation}
 under the  affine  Weyl group symmetry of type $ A^{1}_{l}$.
Here field variables $u$,$q$ and $r$ are matrices and variable $z$ treated as scalar commuting object with field variables, therefore mathematical forms  of classical PII equation and Matrix PII equation seem similar.
   \subsection{Non-commutative PII equation}
 In the context of extension of classical  Painlev\'e equations to non-commutative spaces, a very initial achievement in this  direction was obtained by V. Retakh and V. Roubtsov in \cite{7} where non-commutative analogue of classical Painlev\'e II equtaion obtained through non-abalian Painlev\'e II  symmetric form connected to  noncommutative Toda chain. The non-commutative Painlev\'e II equation derived in folowing form
 
 \begin{equation}\label{VRVR} 
  u^{''} = 2 u^{3}- 2[z,u]_{+} + 4 ( \beta + \frac{1}{2} ), 
 \end{equation} 
 
here $[z,u]_{+}$ is the anti-commutation relation between field variable $u$ and variable $z$ which gives the pure sense of non-commutativity but still here we do not have the explicit commutation relation between field variable $u$ and variable $z$ in non-commutative settings.
\section{Quantum PII  Lax pair}
This section includes the presentation of new Lax pair, Quantum Painlev\'e II  Lax Pair, which directly involves the Planck constant and an arbitrary field variable. In subsequent proposition$2.1$,  it is shown that the compatibility of Quantum Painlev\'e II  Lax Pair simultaneously yields   Matrix Quantum Painlev\'e II equation and quantum commutation relation between field variable $u(z; \alpha) $ and variable $z$. Further it is elaborated with different choices of arbitrary field varible the system reduces to Non-comuutative quantum Painlev\'e II,  derivative matrix  Painlev\'e II equation \cite{OS} and to its classical analogues.   The Quantum Painlev\'e II  Lax Pair incorporates two addition objects which make this new pair different from Flaschka-Newell Lax pair, as the arbitrary field variable $v(z)$ and Planck constant explicitly. The arbitrary field can be chosed in four different ways and the resulting linear system consistants with Painlev\'e II  equation equation under classical limit, briefly  four cases are as   (i) if the arbirary field variable is take as $v=u^{'}$ , we get non-comuutative quantum Painlev\'e II equation with quantum commutation relation $\frac{d}{dz}( zu-uz)=-\frac{i}{2} \hbar  u $ , (ii) for arbitrary field variable $v=u$ the system gives rise to derivative matrix Painlev\'e II  equation with quantum commutation relation $zu-uz=-\frac{i}{2} \hbar u$. , (iii) under the classical limit with $v(z)=u(z; \alpha)$ as scalar we obtain a new classical Painlev\'e II  Lax Pair, one the member  of that pair possesses additional term which makes it different from  Flaschka-Newell  Painlev\'e II  Lax pair and its compatibility consistants with  classical Painlev\'e II equation  (\ref{CPII})
, (iv) under the classical limit with $v(z)=0$ the Quantum Painlev\'e II  Lax Pair reduces to Flaschka-Newell  Painlev\'e II  Lax pair .
\subsection*{Proposition 3.1}
The compatibility  condition of following linear system
\begin{equation} \label{QPL}
    \Psi^{'}=P\Psi,\hspace{1cm}\Psi_\lambda=Q\Psi,
\end{equation}
with matrices
 \begin{equation}\label{QPM} 
\left\{
\begin{array}{lr}
  P=u\sigma_1-i \lambda \sigma_3+4vI \\
    Q=-(4 i \lambda^2+i z+2u^2)\sigma_3+(4 \lambda u-\frac{\alpha}{\lambda})\sigma_1-(2u^{'}-i \hbar)\sigma_2
\end{array}
\right.,
 \end{equation} 
 simultaneously yields 
 \begin{equation}\label{QMPII} 
\left\{
\begin{array}{lr}
   u^{''}=2u^3-\frac{1}{2}[z,u]_{+}+4[v,u^{'}]_{-}+\alpha\\
    zv-vz=-\frac{i}{2} \hbar u
\end{array}
\right.,
 \end{equation} 
here $I$ is identity matrix of order $2$ and $\hbar$ is Planck constant, $v(z)$ is arbitrary field variable.\\
\textbf{Proof:}\\
This can be shown that  from linear  system (\ref{QPL}) we can calculate $  ( \Psi^{'} )_\lambda =   ( \Psi_\lambda)^{'} $  in following form
\begin{equation}\label{ZC4} 
 Q_{z}-P_{\lambda}= [P,Q]_{-},
\end{equation} 
 We can easily  evaluate the values for  $ Q_{z}$, $P_{\lambda}$ and $[P,Q]_{-} = PQ -QP$ from the linear system (\ref{QPL}) as follow
\begin{equation}\label{V1} 
 Q_{z} = -i(  2u^{'} u + 2u u^{'}  +1 )\sigma_{3} -2u^{''} \sigma_{2} + 4 \lambda u^{'}\sigma_{1} ,
\end{equation} 

\begin{equation}\label{V2} 
 P_{\lambda} =  -i \sigma_{3} ,
\end{equation} 
and now
\begin{equation}\label{V3} 
 Q_{z}- P_{\lambda} = \begin{pmatrix}
-2i [u ,u^{'} ]_{+}  & 4 \lambda u^{'} + 2iu^{''}\\
4 \lambda u^{'} -2iu^{''} & 2i [u ,u^{'} ]_{+}
\end{pmatrix} ,
\end{equation}
\begin{equation}\label{V4} 
[ P,Q ]_{-} =   \begin{pmatrix}
 i [z, v]_{-} -2i  [u, u^{'}]_{+}  -\frac{1}{2}\hbar u  &  \delta^{+}  \\
  \delta^{-}  &   -i [z, v]_{-} +2i  [u, u^{'}]_{+}  +\frac{1}{2}\hbar u  
\end{pmatrix},
\end{equation} 

where
\[   \delta^{+} = 4\lambda u^{'}+ 4iu^{3}+i[z,u]_{+} + 2i \alpha  + 2i [v, u^{'}]_{-}- 2i \lambda   \hbar,\]

and 
\[  \delta^{-} = 4\lambda u^{'} -4iu^{3}- i[z,u]_{+} - 2i \alpha  -2i \hbar [v, u^{'}]_{-}- 2i\lambda   \hbar,\]
now  substituting the values of $ L_{z}- P_{\lambda} $ and $[ P,L ]_{-} $from (\ref{V3}) and (\ref{V4}) into zero curvature condition equation  (\ref{ZC4}) and after some simplification, then equating the corresponding elements of resulting matrices on both side, we get
\begin{equation}\label{L4} 
 [ z,v] = -\frac{1}{2}i  \hbar u,
\end{equation} 
and
\begin{equation}\label{L5} 
u^{''} = 2u^{3} -\frac{1}{2}[z,u]_{+} +  \alpha  + [v, u^{'}]_{-}- \lambda   \hbar,
\end{equation} 
\begin{equation}\label{L6} 
u^{''} = 2u^{3}-\frac{1}{2}[z,u]_{+} +  \alpha  + [v, u^{'}]_{-}+ \lambda   \hbar,
\end{equation} 
Now adding  (\ref{L5}) and (\ref{L6}) we obtain 
\begin{equation}\label{L7} 
u^{''} = 2u^{3}-\frac{1}{2}[z,u]_{+}   + [v, u^{'}]_{-} +  \alpha,
\end{equation} 
\subsection{Case-i}  {Taking $v=u^{'}$ }\\
With the choice of $v=u^{'}$ Quantum Matrix Painlev\'e II  (\ref{QMPII})  reduces to the following form 
 \begin{equation}\label{QPII} 
\left\{
\begin{array}{lr}
   u^{''}=2u^3-\frac{1}{2}[z,u]_{+}+\alpha \\
\frac{d}{dz}( zu-uz)=-\frac{i}{2} \hbar  u 
\end{array}
\right.,
 \end{equation} 
 the last expression in above equation (\ref{QPII})  shows the quantum commutation relation between independent variable $z$ and field variable $u$. 
\subsection{Case-ii} {Matrix field $v=u$  and derivative matrix PII equation:}\\
Taking  derivation  of Quantum matrix PII equation  (\ref{QMPII}) with respect to $z$, we gat  
\begin{equation}
    u^{'''}=(2u^3)^{'}- \frac{1}{2}[2uz-\frac{i}{2}\hbar u]^{'}+4[u,u^{'}]_{-}^{'},
\end{equation}
\begin{equation}
    u^{'''}=2u^2 u^{'}+2u^{'}u^2+2uu^{'}u-\frac{1}{2}[2u+2zu^{'}-\frac{i}{2}\hbar u^{'}]+4[u,u^{''}]_{-},
\end{equation}
or
\begin{equation}
    u^{'''}=2u^2u^{'}+2u^{'}u^2+2uu^{'}u-u-(z-\frac{i}{4}\hbar )u^{'}+4[u,u^{''}]_{-},
\end{equation}
now introducing  new field variable $\nu(x)=u(z)$  where  $x=z-\frac{i}{4}\hbar $  in above expression, we obtain
\begin{equation}
    \nu^{'''}=2\nu^2\nu^{'}+2\nu^{'}\nu^2+2\nu \nu^{'}\nu-\nu-x\nu^{'}+4[\nu,\nu^{''}]_{-},
\end{equation}
above equation is not exactly but  similar to derivative matrix Painlev\'e II equation \cite{OS} obtained dimensional reduction of matrix mKdV equation which differs by two additional terms  $2\nu^2\nu^{'}+2\nu^{'}\nu^2$ but under the classical limit both coincide. \subsection{Case-iii} {under the classical limit as  $\hbar \rightarrow 0 $: }\\
Under the classical limit  $\hbar \rightarrow 0 $ as  the qunatum commutation relation vanishes and above system (\ref{QMPII})  reduces to its classical analogue  and compatibility condition  of pair  (\ref{QPM}) under this limit still consistants for the classical Painlev\'e II  equation,  where as the additional term $v=u^{'}$ at diagonal of $P$ makes that pair   different  from  Flschka-Newell Lax pair in classical case, if we take $v=0$ that pair exactly reduces to    Flschka-Newell Lax Pair. Here this has been demonstrated  that Flaschk-Newell pair appears as  case of our newly presented  Quantum Painlev\'e II Lax pair  (\ref{QPM}).  
  \section{Gauge Equivalence of Quantum PII Lax Pair }  \subsubsection{\textbf{Proposition 4.1}}
 The compatibility of gauge equivalent  Quantum  PII Lax pair  $\tilde{P}= GPG^{-1}$ and $\tilde{Q}= GQG^{-1}$ 
 \begin{equation}\label{PQT} 
\left\{
\begin{array}{lr}
\tilde{P}= u \sigma_3 -i \lambda  \sigma_2+4u I\\
\tilde{Q}= (4 \lambda u-\frac{\alpha}{\lambda})\sigma_3 -(4i \lambda^2+ \frac{1}{4}\hbar)\sigma_2+2pI_{+} - 2qI_{-} 
\end{array}
\right.,
 \end{equation} 
 yields quantum non-ablian  set of three equation
      \begin{equation}\label{qspII} 
\left\{
\begin{array}{lr}
 p^{'}= vp-pv+ up+ pu - i\frac{1}{4}\hbar u  - \alpha+\frac{1}{2} \\
 q^{'}= qv-vq-uq -qu - i\frac{1}{4}\hbar u  + \alpha+\frac{1}{2} \\\\
u^{'}= \frac{1}{2}( p -q)
\end{array}
\right.,
 \end{equation} 
 where  $p=u^2+ u^{'}+ \frac{z}{2}$ , $q=u^2- u^{'}+ \frac{z}{2}$ and   $ G= \frac{1}{ \sqrt{2}}\begin{pmatrix} 
-i & -i \\
-1 & 1
\end{pmatrix}$, 
  $ I_{+}= \begin{pmatrix}
0 & 1 \\
0 & 0
\end{pmatrix} $,  $I_{-}= \begin{pmatrix}
0 & 0 \\
-1 & 0
\end{pmatrix}. $\\
\textbf{Proof:}\\
It is  straight forward to construct $\tilde{P}$ and $\tilde{Q}$ from  (\ref{QPM}) under  gauge transformations   $\tilde{P}= GPG^{-1}$ and $\tilde{Q}= GQG^{-1}$.\\

  The compatibility of linear system $\Psi^{'}=\tilde{P} \Psi,\hspace{1cm}\Psi_\lambda= \tilde{Q} \Psi$  gives rise to  the zero-curvature condition as $\tilde{Q} -\tilde{P}_{\lambda}= [\tilde{P},\tilde{Q} ]_{-}$   which implies with help of matrices (\ref{PQT}) to get the set of equations (\ref{qspII}) as the quantum non-abelain analogue of  (\ref{csp2}).\\
 Now the first equation from system  (\ref{qspII}) with arbitrary field $v=u^{'}$ can be written 
 \begin{equation}\label{qpII} 
p^{'}= 2up - i\frac{1}{4}\hbar u  - \alpha+\frac{1}{2}, \\
 \end{equation} 
or 
 \begin{equation}\label{fqspII} 
p^{'}= 2u(p - \frac{\beta}{2} )  - \delta ,
 \end{equation}
 here $\beta =  i\frac{1}{4}\hbar $, $\delta= \alpha-\frac{1}{2}$ and if we take $\mathbf{ p} =p - \beta$ then above expression will take the following form
  \begin{equation}\label{fqpII} 
u= \mathbf{p^{'} }\mathbf{p^{-1} } + \delta \mathbf{p^{-1} } ,
 \end{equation}
 with 
    \begin{equation}\label{fp} 
    \mathbf{p}=  u^{2} +  u^{'}+\frac{z}{2} - \frac{\beta}{2}, 
 \end{equation}
   
 It is straight forward to calculate  $u^{'}$ and  $u^{2}$  from expression (\ref{fqpII}) as below 
  \begin{equation}\label{uup} 
\left\{
\begin{array}{lr}
 u^{'}= -\frac{1}{2}\mathbf{p^{'} }\mathbf{p^{-1} }\mathbf{p^{'} }\mathbf{p^{-1} }+ \frac{\delta }{2}\mathbf{p^{-1} }\mathbf{p^{'} }\mathbf{p^{-1} }\\
  u^{2}= \frac{1}{4}\mathbf{p^{'} }\mathbf{p^{-1} }\mathbf{p^{'} }\mathbf{p^{-1} }+ \frac{\delta }{4}\mathbf{p^{'} }\mathbf{p^{-2} }+\frac{\delta }{4}\mathbf{p^{-1} }\mathbf{p^{'} }\mathbf{p^{-1} }+\frac{\delta^{2} }{4}\mathbf{p^{-2} }
 \end{array}
\right.,
 \end{equation} 
 Now substituting the values of $u^{'}$ and  $u^{2}$ into  expression  (\ref{fp}) and then after doing some simplification, we get
  \begin{equation}\label{uup34} 
 \mathbf{p}^{''} = -\frac{1}{2}\mathbf{p^{'} }\mathbf{p^{-1} }\mathbf{p^{'} } + 2\mathbf{p^{2} }- \frac{\delta^{2} }{2}\mathbf{p^{-1} } - (z-\beta ) \mathbf{p } ,
   \end{equation} 
 Here during simplification  the non-commutative definition of  logarithmic derivative  \cite{7}  is  used as  $\frac{d}{dz} ln \mathbf{p}= \mathbf{p^{'} }\mathbf{p^{-1} } $ or $\frac{d}{dz} ln \mathbf{p}= \mathbf{p^{-1} } \mathbf{p^{'} }$.
The above can be regraded as Non-abelian quantum P34 equation for  $\mathbf{p}$ which involves Planck constant with power $+1$ as $ \hbar $ which rescues its to be negligible as compare to  $ \hbar^{2} $  where as in  \cite{NH, NGR}  quantum P34 incorporates $ \hbar^{2} $  that is much smaller   then $ \hbar $  and  can be assumed  negligible as compare to $ \hbar $. Therefore the presence of Planck constant as $ \hbar $ in  (\ref{uup34}) consolidates   P34 equation from quantum mechanical point of view as compare to the P34 equation possesses  Planck constant with higher positive powers as  $ \hbar^{2} $.   \\
It is straight  forward to see that under the classical limit as  $\hbar \rightarrow 0 $ quantum  P34 equation  (\ref{uup34})  reduces to its classical analogue obtained from  system (\ref{csp2}) which arises from compatibility  Flaschka-Newell  gauge equivalent Lax pair. Similarly  for variable $q$ we obtain
  \begin{equation}\label{uuq34} 
 \mathbf{q}^{''} = -\frac{1}{2}\mathbf{q^{'} }\mathbf{q^{-1} }\mathbf{q^{'} } + 2\mathbf{q^{2} }- \frac{\delta^{2} }{2}\mathbf{q^{-1} } - (z+\beta ) \mathbf{q }.
   \end{equation} 

 \subsubsection*{Remark 4.1}This also can be shown that if we eliminate $p$ and $q$ from   (\ref{qspII})  the resulting equation will non-commutative PII equation. For this purpose take derivation of last equation of system  (\ref{qspII})  as   $u^{''}= \frac{1}{2}( p^{'} -q^{'})$ then use the values  $p^{'}$ and $q^{'}$ and after that substituting the values of $p=u^2+ u^{'}+ \frac{z}{2}$ , $q=u^2- u^{'}+ \frac{z}{2}$ in  resulting expression we get non-commutative PII equation  \cite{7}.
 \section{Conclusion}
In this work we presented a  quantum Painlev\'e II  Lax Pair which directly involves the Planck constant $\hbar$ and an arbitrary field variable $v$ whose  compatibility condition  is equivalent to the quantum Painlev\'e II equation and a quantum commutation relation between field variable $v $  and   independent  variable $z$ simultaneously which  it more physical from quantum mechanical point of view. We have also shown that its gauge equivalent pair   gives rise to non-abelian PII symmetric form involving Planck constant $\hbar$ further is reduced to  quantum p34 equations with non-commutative PII equation and we made a substantial  comparison of our results to existed.  It has been manifested that  all our calculated results coincide to their classical analogues under $\hbar \rightarrow 0 $ for scalar case.  For further motivation, it seems quite interesting to construct the quantum matrix analogue of classical mKdV equation $\And$ KdV equation from presented quantum p34 equation through reverse scale transformations. More interestingly to investigate the pure quantum analogue of nonlinear system of equations associated to Toda chain at $n=1$ with quantum commutation relations  and which further  can be applied to obtain some new results for higher values of $n$ with the help of  quantum Painlev\'e II setting presented in this paper towards the quantization of  Painlev\'e II  equation with its connected integrable systems.

 \section*{Acknowledgement}
This research work has been completed as a part of Belt and Road Young Scientist Plan—Science and Technology Innovation Plan of Shanghai, China (No. 20590742900), sponsored by Science and Technology Commission of Shanghai Municipality, China on providing us the facilities to complete this work. I am thankful to Shanghai University on providing me facilities and also to the Punjab University 54590, Lahore for the partial financial support to complete this project at Shanghai University, China. The partial results of this project are completed during my stay at BIMSA, China, my sincere thanks to the organisers of Representation Theory, Integrable Systems and Related Topics-Attendance and BIMSA for their hospitality and facilities.

%\bibliography{<your-bib-database>}

\end{document}